\documentclass[%
 amsmath,amssymb,
 preprint,%
nofootinbib,
]{revtex4-1}
\pdfoutput=1
\usepackage{rotating} 
\usepackage{graphicx,epsfig}

\usepackage{amsmath}
\usepackage{amssymb}
\usepackage{physics}
\usepackage{subfigure}
\usepackage{color}
\usepackage{mathtools}
\usepackage{tensor}

\usepackage{tabularx}
\usepackage{dcolumn}
\usepackage{multirow}

\usepackage{appendix}
\usepackage{relsize}

\usepackage[colorlinks=true
  ,urlcolor=blue
  ,anchorcolor=blue
  ,citecolor=blue
  ,filecolor=blue
  ,linkcolor=blue
  ,menucolor=blue
  ,linktocpage=true
  ,pdfproducer=medialab
  ,pdfa=true
]{hyperref}

\def\beq{\begin{equation}}
\def\eeq{\end{equation}}
\def\bea{\begin{eqnarray}}
\def\eea{\end{eqnarray}}
\def\be{\begin{equation}}
\def\ee{\end{equation}}

\def\bse{\begin{subequations}}
	\def\ese{\end{subequations}}

\def\Mp{M_{\rm pl}}
\def\l{\left}
\def\r{\right}
\def\wre{w_{\rm re}}
\def\nre{N_{\rm re}}
\def\tre{T_{\rm re}}
\def\ns{n_{_{\mathrm{s}}}}
\def\d{\mathrm{d}}

\graphicspath{{./figs/}}

\begin{document}
	
	\title{Model-independent constraints on inflation and reheating}
	
	\author{Pankaj~Saha}
 \email{pankaj@physics.iitm.ac.in}
\affiliation{ 
Centre for Strings, Gravitation and Cosmology, Department of Physics, Indian Institute of Technology Madras, Chennai, Tamil Nadu 600036, India
}%

\begin{abstract}
Reheating connects the inflationary universe to the radiation-dominated evolution of standard Big Bang cosmology. Due to the lack of direct observations, we rely on indirect bounds on this phase from the Cosmic Microwave Background~(CMB) data. Using reheating constraints to arrive at additional constraints on inflationary models is prevalent in the literature. In this work, we develop a formalism to analyze the reheating constraints for the general case of single field canonical slow-roll inflation without dealing with a specific inflationary model. We find that using the lower bound on reheating temperature and the upper bound on the tensor-to-scalar ratio: one can constrain the inflationary energy scale and the duration of the reheating epoch after slow-roll inflation. Following the standard practice, we described the reheating phase with an effective equation of state parameter~($w_{\rm re}$). However, with the present formalism, we can have quantitative information of the reheating phase even when the reheating equation of state behaves as radiation $w_{\rm re} = 1/3$. For the canonical reheating phase described by~$w_{\rm re}<1/3$, we find that the inflationary efolding number must be bounded from above, i.e., $N_k\leq56.09$. Consequently, if inflationary efolds satisfy this upper bound \textemdash i.e., $N_k=56.09$ \textemdash a finite period of reheating is only possible for exotic reheating phases described by $w_{\rm re}>1/3$. We also find that as we lower the value of $N_k$, the corresponding energy scale of inflation also decreases. Consequently, it will be more and more difficult to detect the corresponding inflationary gravitational wave signals with the Gravitational Waves mission. Finally, we show how an extended period of canonical reheating can improve the situation in those cases.
\end{abstract}

 
 \maketitle
\tableofcontents
\section{\label{sec:level1}Introduction}
The quantum fluctuations generated during inflation~\cite{Brout:1977ix,Sato:1980yn,Guth:1980zm,Linde:1981mu,Starobinsky:1982ee} account for the anisotropic patterns in the CMB. Inflation is usually studied in a model-dependent way where a scalar field displaced from its true vacuum provides the energy density for the quasi-exponential expansion of the universe. The classification of inflationary models, as a consequence, is often synonymous with the corresponding scalar field potentials. Furthermore, the study of single field canonical slow-roll inflation and their perturbations provide us with various model-independent results (see, for instance,~\cite{Liddle:1993fq,Lyth:1996im,Creminelli:2004yq,Weinberg:2008zzc}). The perturbations help us define observables such that the connection could be made between theory and cosmological observations. For instance, the inflationary scalar power spectrum~($\mathcal{P}_{\mathcal{R}}$) and the spectral tilt~($\ns$) has been precisely measured at large scales~\cite{WMAP:2003syu,Akrami:2018odb}. The tensor mode with its associated power spectrum~($\mathcal{P}_t$) is usually parameterized by the so-called tensor-to-scalar ratio~$r_k (=\mathcal{P}_{t}/\mathcal{P}_{\mathcal{R}})$ whose value is bounded from above. The detection of tensor mode is considered the smoking gun for the inflationary paradigm that will unequivocally determine the scale of inflation. From the perspective of high energy physics, it will ascertain the energy scale which was operative at the earliest phase of the universe amenable within Einstein's theory of gravitation and a minimal extension of the standard model of particle physics~\cite{Lyth:1998xn,Baumann:2009ds,Mazumdar:2010sa,Baumann:2018muz}.
\par
The measurement of the inflationary energy scale can be made directly from the inflationary stochastic background of gravitational waves~(SGWs) with the future detectors such as LISA~\cite{Bartolo:2016ami},~DECIGO~\cite{Kawamura:2008zza},~BBO~\cite{Crowder:2005nr}; or indirectly through its effect on the polarization of the cosmic microwave background radiation~(CMB) with missions such as CMBPol~\cite{CMBPolStudyTeam:2008rgp},~PRISM~\cite{PRISM:2013fvg},~\cite{CORE:2016ymi} which will be able to probe $r_k$ at $10^{-3}$ level. However, it has been argued that measuring $r < 10^{-4}$ via CMB polarization can be highly challenging~\cite{Knox:2002pe} while cosmic variance limits it at a level of $10^{-5}$. Additionally, some prospects of determining the energy scale of inflation were presented in~\cite{Bartolo:2018qqn} by probing the details of primordial non-Gaussinities contained in the large-scale structure surveys as DESI~\cite{DESI:2016fyo},~LSST~\cite{LSSTScience:2009jmu}, and Euclid~\cite{EUCLID:2011zbd} while determining the scale within the context of Feebly Interacting Massive Particle (FIMP) model of Dark Matter~(DM) was studied in~\cite{Enqvist:2017kzh}. In this work, we will describe a way to constrain the energy scale of inflation and the inflationary efolding using the constraints from reheating in a model-independent approach.
\par
After the end of inflation, the inflaton starts oscillating around its true vacuum and decays into other matter fields and radiation depending on the coupling to such fields. The result of the process is to give way to a thermalized radiation-dominated universe at around 10 MeV so as not to ruin the successful predictions of the standard big bang nucleosynthesis~(BBN)~\cite{Kawasaki:1999na,Steigman:2007xt}. This intermediate phase in the evolution of the universe between the end of inflation to the commencement of a radiation-dominated universe is known as the reheating phase. Thus reheating, which defrosts the universe after the colossal amount of expansion during inflation, is the true progenitor of the Big bang cosmology. Despite its importance, the phase of reheating is perhaps the least understood phase in the early universe. The difficulty is partly due to the lack of direct observables. In addition, the process of thermalization erases much of the information of the phase. In light of these limitations, the reheating constraints on inflationary models\textemdash tracking the expansion of the universe and the dilution of the energy density from inflation to the present epoch\textemdash is generally described with the following two quantities: the duration of the phase\textemdash the reheating efolding number~$N_{\rm re}$ and the temperature of the universe at the end of the phase\textemdash the reheating temperature~$T_{\rm re}$. The minimum value of reheating efolding number, $N_{\rm re} = 0$, corresponds to instantaneous reheating; however, there are no model-independent bound on the maximum duration of reheating. On the other hand, the reheating temperature is bounded from above, for a specific inflationary model, by the inflationary energy scale, which for instantaneous thermalization can be directly used to find the maximum reheating temperature. The minimum reheating temperature, as we just mentioned, is set by the condition of successful BBN.
\par
The reheating constraints on inflationary models have been studied extensively in the literature in recent times\cite{Mielczarek:2010ag,Dai:2014jja,Martin:2014nya,Munoz:2014eqa,Cook:2015vqa,Martin:2016iqo,Ji:2019gfy}. However, a model-independent study of reheating constraints analysis on the inflationary phase is absent in the literature. In this work, we will use the bound on the inflationary tensor-to-scalar ratio and the minimum reheating temperature required for BBN to constrain the inflationary and reheating parameters. We will not consider any specific form of the potential; hence our results allow us to arrive at bounds on the inflationary phase in a model-independent way. Although a single component dominates the dynamics during inflation, the reheating phase is described by multiple components. Thankfully, the system can be economically described with an effective fluid parameterized by an equation of state parameter~(EoS)~$w_{\rm re}$ as is well known in the literature~\cite{Dai:2014jja,Martin:2014nya,Saha:2020bis}. Our primary motivation in this work is to constrain the inflationary phase viz. the inflationary energy scale~(the tensor-to-scalar ratio $r_k$) at the Planck pivot scale from reheating constraints. In doing so, we find that there exists an upper bound on inflationary efolding ($N_k\leq56.09$) for the canonical reheating phase described by a non-stiff equation of state parameter~($\wre<1/3$). Considering a single field canonical slow-roll inflation, we have determined the energy scale of inflation and show how the scale varies with the duration of reheating and the inflationary efolding number. We consider the variations for different EoS. Our formalism enables us to find the allowed duration of reheating and the consequent variation of the inflationary energy scale even when the reheating EoS is radiation-like ($\wre=1/3$).
\par
The remainder of the paper is structured as follows: After describing the basic equations, and the connections between reheating epoch to observables in~Sec~\ref{sec:sec2}, We presented our results in Sec \ref{sec:sec3}. We conclude in Sec. \ref{sec:conc}. We have used the usual Friedmann-Lema\^itre-Robertson-Walker~(FLRW) metric $ds^2 = -\d{t}^2 + a(t)^2(\d{x}^2+\d{y}^2+\d{z}^2)$, with $a(t)$ being the scale factor and $H=\dot{a}/a$ is the Hubble expansion rate with the overdot being the derivative with respect to time. We denote $N=\ln(a)$ as the efolding number measuring the expansion of the universe. We take $\Mp (=2.43\times10^{18}~\textrm{GeV})$ to denote the reduced Planck mass. 

\section{\label{sec:sec2}Reheating}
This section will present all the required relations for connecting the reheating and inflationary phase with present CMB parameters. We will then describe how to relate the Hubble scale at the end of inflation to that instant when a mode exits the horizon. Equipped with these two relations, we will analyze the phases without dealing with a specific form of the inflationary potential.
\subsection{\label{sec:sec2o1}Reheating constraints on inflationary phase}
\begin{figure}[t]
\centering
 \includegraphics[width=\textwidth]{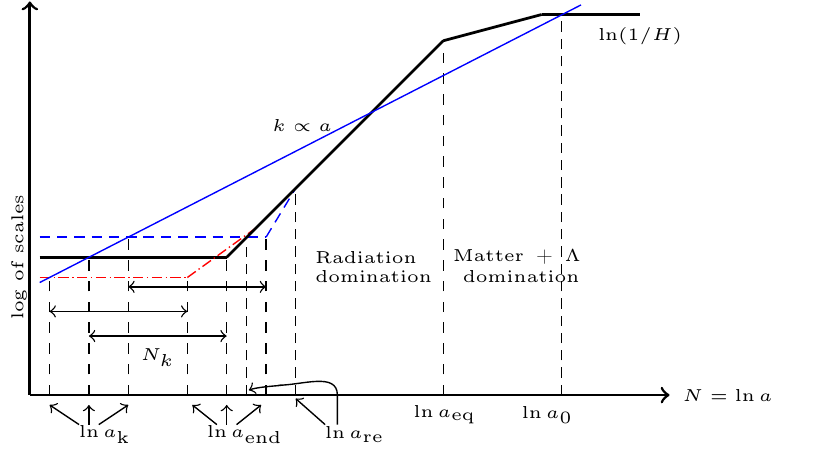}
 \caption{\scriptsize A sketch of the variation of inflationary energy scale with reheating efolding numbers for different EoS. The solid black line shows the evolution of the Hubble horizon~($1/H$) at different epochs when the reheating process is instantaneous. The red dash-dotted line is with reheating with EoS parameter $\wre<1/3$ while the blue dashed line when the reheating EoS parameter is $\wre>1/3$. The solid blue line is any reference physical perturbation mode that leaves the horizon at $N_k$ efolds before the end of inflation. The redshift at the matter-radiation equality is precisely known in terms of the present matter abundance and the value of the Hubble parameter today. The level of anisotropy in the CMB can be explained if inflation lasts for 50-60 efolds~\cite{Mukhanov:2005sc,Remmen:2014mia}. For a fixed $N_k$, an increase in reheating epoch will increase the inflationary energy scale $H_k$ when reheating EoS is $\wre<1/3$ while the opposite happens if reheating is described by an EoS $\wre<1/3$. It is also easy to see that for any mode, increasing $N_k$ should consequently increase the inflationary energy scale.}
 \label{fig:sch_hubble}
\end{figure}

The Einstein equations describing the background dynamics in the usual FLRW spacetime is
\begin{align}
\label{ein1}
3 \Mp^2\,H^2 &= \rho,\\
\label{ein2}
2 \Mp^2\,\dot{H} &= -\left(\rho + P\right),  
\end{align}
where, $\rho$ and $P$ are, respectively, the total energy and pressure of the components describing the system. The phase of inflation is described by a phase of shrinking Hubble radius:
\begin{equation}
 \dv{t}(aH)^{-1} < 0,
 \label{eq:dhr}
\end{equation}
while during any other phase of cosmic evolution the Hubble radius increases (cf. Fig.(\ref{fig:sch_hubble})). The condition of inflation in (\ref{eq:dhr}) can also be written as:
\begin{align}
 -\frac{1}{a}(1-\epsilon) &< 1,\\
 \implies \epsilon &< 1,
\end{align}
where $\epsilon\equiv-\dot{H}/H^2$ is known as the first slow-roll parameter. Clearly, inflation ends when $\epsilon=1$.
\par
Writing the instanteneous equation of state parameter as $w(N)=p(N)/\rho(N)$, the conservation of stress-energy tensor $\nabla_{\mu}\tensor{T}{^\mu_\nu}=0$ imples that the energy density varies with the expansion as 
\begin{equation}
 \rho(N) = \rho_{\rm end} \exp\Bigg(-3\int_0^{N}\big(1+w(N')\big)dN'\Bigg).
\end{equation}
Now, defining the effective EoS parameter $w_{\rm re}$ during reheating as:
\begin{equation}
 w_{\rm re} = \frac{1}{N_{\rm re}}\int_{0}^{N_{\rm re}}w_{\rm eff}(N'){\rm d}N'.
\end{equation}
we have the energy density at the end of reheating
\begin{equation}
 \rho_{\rm re} \equiv \rho(N)\Big|_{N=N_{\rm re}} = \rho_{\rm end} \exp\Big(-3(1+w_{\rm re})N_{\rm re}\Big).
\end{equation}
The duration of reheating epoch or the reheating e-folding number $N_{\rm re}$ can, thus, be written as
\begin{eqnarray}\label{Eq:NreRho}
	N_{\rm re} = \frac{1}{3(1 + w_{\rm re})}\ln \left(\frac{\rho_{\rm end}}{\rho_{\rm re}}\right).
\end{eqnarray}
Now, using the relation $k = a_k H_k$, i.e., the perturbation mode that we observe today exited the comoving horizon at an efolding $N_k$ before the end of inflation, we arrive at
\begin{align}
	\frac{k}{a_0 H_0} &= \frac{a_k}{a_{\rm end}} \frac{a_{\rm end}}{a_{\rm re}}\frac{a_{\rm re}}{a_{\rm eq}} \frac{a_{\rm eq} H_{\rm eq}}{a_0 H_0}\frac{H_k}{H_{\rm eq}} \nonumber\\
	\label{Eq:ka0}
	\implies\ln \left(\frac{k}{a_0 H_0}\right) &= -N_k - N_{\rm re} - N_{\rm RD} + \ln \left(\frac{a_{\rm eq} H_{\rm eq}}{a_0 H_0}\right)  +\ln \left(\frac{H_k}{H_{\rm eq}}\right),
\end{align}
where $N_{\rm RD} \equiv \ln(a_{\rm eq}/a_{\rm re})$. The subscripts `$0$' and `eq' represent the parameters at present epoch and the epoch of matter-radiation equality, respectively. 
Assuming a complete conversion of inflaton energy into radiation at the end of reheating, the energy density at the end of reheating is related to the reheating temperature ($T_{\rm re}$) as: $\rho_{\rm re} = \rho_{\rm rad}= (\pi^2/30) \,g_{\rm re} \,T_{\rm re}^4$, where $g_{\rm re}$ accounts for the effective number of relativistic species upon thermalization. As most of the entropy production in the universe occurs during the reheating phase, we can assume that the reheating entropy is preserved in the present CMB and the neutrino background. This helps us to connect the reheating temperature with the present CMB temperature as
 	\begin{eqnarray}\label{Eq:TreT0}
 		\frac{T_{\rm re}}{T_0} = \left(\frac{43}{11 g_{\rm s, re}}\right)^{1/3}\,\frac{a_0 }{a_{\rm eq}}\frac{a_{\rm eq}}{a_{\rm re}},
 	\end{eqnarray}
 	\noindent where, we have used the relation between the background neutrino temperature and the CMB temperature $T_{\nu0} = (4/11)^{1/3}\,T_0$ and $g_{\rm s, re}$ is the effective number of light species for entropy at reheating. Substituting this relation in \eqref{Eq:NreRho}, we obtain
\begin{align}
\frac{3\l(1 + w_{\rm re}\r)}{4}\, N_{\rm re} = \frac{1}{4} \ln \left(\frac{30}{g_{\rm re} \pi^2}\right) + \frac{1}{4} \ln \left(\frac{\rho_{\rm end}}{T_0^4}\right) + \frac{1}{3} \ln \left(\frac{11 g_{\rm s, re}}{43}\right) + \ln \left(\frac{a_{\rm eq}}{a_0}\right) - N_{\rm RD}.
\end{align}
Combining this relation with \eqref{Eq:ka0}, the reheating e-folding number can be re-written as \cite{Liddle:2003as}
\begin{align}
\nonumber
N_{\mathrm{re}} = \frac{4}{ 3 w_{\mathrm{re}} - 1}\Bigg[ N_{k} - \ln\left(\frac{H_k}{\Mp}\right) +\frac{1}{4}\ln\l(\frac{\rho_{\rm end}}{\Mp^4}\r) +\ln \left(\frac{k}{a_{0} T_{0}}\right)  + \frac{1}{4} \ln \left(\frac{30}{\pi^{2} g_{\rm re}}\right)  \\
+ \frac{1}{3}\ln \left(\frac{11 g_{\rm s, re}}{43}\right)\Bigg].
\end{align}
Using the Friedmann equation, we can write.
\begin{align}
 \ln\left(\frac{H_{\rm end}}{H_k}\frac{\Mp}{H_k}\right) = \frac{1}{2}\Bigg[(3w_{\rm re}-1)N_{\rm re} -4N_k -
 \ln\l(\frac{90}{\pi^2 g_{\rm re}}\r) - \frac{4}{3}\ln\l(\frac{11g_{\rm re}}{43}\r) - 4\ln\l(\frac{k}{a_0 T_0}\r)\Bigg].
\end{align}
Substituting the numerical values of $T_0=2.723\mathrm{K}$ and considering a Minimal Supersymmetric extension of the Standard Model of particle physics~(MSSN), we can take ${\{g_{\rm re}, g_{\rm s, re}\}}\sim 100$, we have
\begin{align}
 \ln\l(\frac{H_{\rm end}}{H_k}\frac{\Mp}{H_k}\r) = \frac{1}{2}\Bigg[(3w_{\rm re}-1)N_{\rm re} -4(N_k+29.56) -4\ln k\Bigg].
 \label{eq:relreh}
\end{align}
Until now, we have not made any assumptions about the inflation model or the nature of inflation. Next will consider a single field slow-roll inflation to connect the energy scales $H_{k}$ to $H_{\rm end}$.
\subsection{\label{sec:sec2o2}Connecting \texorpdfstring{$H_k$}{Hk} to \texorpdfstring{$H_{\rm end}$}{Hend}}
We will follow~\cite{Liddle:1993ch} to connect the inflationary energy scale $H_k$ when the perturbation mode $k$ exit the horizon~($k=a_kH_k$) to the inflationary energy scale at the end of inflation $H_{\rm end}$. Using the Hamilton-Jacobi equation for inflation\cite{Liddle:1993ch}, we can write. 
\begin{equation}
 N_k = \int_{\phi_k}^{\phi_{\rm end}}\frac{1}{\sqrt{2\epsilon(\phi)}}\frac{d\phi}{\Mp},
 \label{eq:nk_hj}
\end{equation}
\begin{equation}
 \frac{H_{\rm end}}{H_k} = \exp\l(-\int_{\phi_k}^{\phi_{\rm end}} \sqrt{\epsilon(\phi)}\frac{d\phi}{\Mp}\r),
 \label{eq:int_hk}
\end{equation}
where at the end of inflation $\epsilon(\phi_{\rm end})\equiv 1$. If we consider a very flat potential such that $\epsilon(\phi)$ is monotonic and $\epsilon'(\phi)\sim 0$ for most of the field space. Then, due to the small value of the parameter $\epsilon(\phi)$, the integral in \eqref{eq:nk_hj} gets most of its contribution around $\phi_k$, we may approximate the integration in (\ref{eq:int_hk}) as 
\begin{equation}
 \frac{H_{\rm end}}{H_k} = \exp\l(-N_k\epsilon_k\r),
 \label{eq:hk_hend_0}
\end{equation}
where we have defined $\epsilon_k\equiv\epsilon(\phi_k)$. We should note that relation~(\ref{eq:hk_hend_0}) is arrived assuming a very flat potential. These flat plateau types of potentials are theoretically well motivated~\cite{Kallosh:2013hoa,Kallosh:2013yoa,Galante:2014ifa,Linde:2015uga,Maity:2019ltu} and are currently favored observationally~\cite{Akrami:2018odb}. Nonetheless, the effect of any variation will be logarithmic and can be taken care of easily. In appendix~\ref{app:appA}, we will examine the variations to this ratio by considering some models.
\par\noindent
Now, for canonical slow-roll inflation, the scalar and tensor spectra are written in terms of $H_k$ as
\begin{align}
\label{eq:pps}
\mathcal{P}_{\mathcal{R}} &= \frac{1}{2\Mp^2\epsilon_k}\left(\frac{H_k}{2\pi}\right)^2\left(\frac{k}{a_k H_k}\right)^{\ns-1},\\
\label{eq:ppt}
\mathcal{P}_{t} &= \frac{8}{\Mp^2}\left(\frac{H_k}{2\pi}\right)^2\left(\frac{k}{a_k H_k}\right)^{n_t}.
\end{align}
In the absence of the detection of the tensor amplitude, the energy scale can thus be fixed from the WMAP value of the amplitude of scalar power spectra modulo the first slow-roll parameter for single-field inflation with a canonical scalar field. Further, using $r_k = 16\epsilon_k$, we can write Eq.(\ref{eq:hk_hend_0}):
\begin{equation}
 \frac{H_{\rm end}}{H_k} = \exp\left(-N_k\frac{r_k}{16}\right).
 \label{eq:hk_hend}
\end{equation}
Further, using Eqs.~(\ref{eq:pps}-\ref{eq:ppt}), the inflationary energy scale for horizon exit of the pivot scale can be expressed in terms of the tensor-to-scalar ratio, and the amplitude of scalar power-spectrum~$(A_s)$ at the pivot scale as
\begin{eqnarray}\label{Eq:HdiffObs}
    H_k = \pi M_{\rm p}\sqrt{\frac{r_kA_s}{2}}. 
    \label{eq:res_hk_rk}
\end{eqnarray}
We notice that the ratio of the Hubble scales defined in~(\ref{eq:hk_hend}) is independent of inflationary models and, together with~(\ref{eq:res_hk_rk}), when substituted in~(\ref{eq:relreh}), will provide us the relation to connecting reheating with inflationary parameters in a model-independent way. 
In our calculation we use the PLANCK~\cite{Aghanim:2018eyx,Akrami:2018odb} pivot scale $k/a_0  = 0.05\, \text{Mpc}^{-1}$. 
Combining the above, We find:
\begin{equation}
 r_k = \frac{W_0(\mathcal{A}\exp(\mathcal{B}))}{\mathcal{A}}.
 \label{eq:res_rk}
\end{equation}
where $W_0$ is the Lambert W function\cite{corless1996lambertw}\footnote{In \texttt{Wolfram~Mathematica}, it is defined as \texttt{ProductLog}~(with a function alias \texttt{LambertW})~\cite{mathmaticalw}} and
\begin{align}
 \mathcal{A} &= \frac{N_k}{8}.\\
 \mathcal{B} &= \ln\l(\frac{2}{\pi^2 A_s}\r) + 4(N_k+29.56+\ln k) + (1-3\wre)\nre.
\end{align}
We will use Eq.(\ref{eq:res_rk}) to find how $r_k$ varies with $\nre$ and $N_k$ which further using Eq.(\ref{eq:res_hk_rk}) can be cast as how energy scale varies. At this point, however, it will be useful to have a diagrammatic view of how the energy scales on inflation varies with $\nre$ and $N_k$ to get an intuitive picture of what to expect from the above relations. We will discuss these dependencies of scales in the next section.
\subsection{\label{sec:sec2o3}How \texorpdfstring{$H_k$}{Hk} depends on \texorpdfstring{$N_{\rm re}$}{Nre} for different EoS}
We can obtain an intuitive picture of the dependence of the inflationary energy scale on $\nre$ and $N_k$ following the evolution of the Hubble horizon at different cosmological epochs as shown in Fig. \ref{fig:sch_hubble}. The solid black line is the evolution of the Hubble horizon when the reheating is instantaneous. The red dashed line is for the case when there is a finite period of reheating with $\wre<1/3$, while the blue dashed line is when $\wre>1/3$. The solid blue line is a reference perturbation mode~$k$ that leaves the horizon $N_k$ efolding before the end of inflation. The reference pivot scale quoted in Planck's results roughly corresponds to the middle of the logarithmic range of scales probed by Planck. For instance, the pivot scale $k_0=0.05~{\rm Mpc^{-1}}$ corresponds to a physical mode of a perturbation of $20~{\rm Mpc}$ that entered the horizon very recently. Now the epoch of matter-radiation equality $(a_{\rm eq})$ counting from the present time $(a_0)$ is precisely known within the measured error of matter abundance~($\Omega_m$) and current Hubble expansion rate~($H_0$). Hence, variation in the evolution history must be compensated by variation in $\nre$ and $N_k$. Now, if we track any reference mode from horizon exit to reentry, it is easy to see from Fig.(\ref{fig:sch_hubble}) that for reheating with $\wre<1/3$, an increase in $\nre$, keeping $N_k$ fixed will increase $H_k$ while the opposite will happen for $\wre>1/3$. For the particular case of $\wre=1/3$, we see that the variation of reheating efolding number will not change the energy scale as the slope of the reheating phase is indistinguishable from the later radiation dominated phase. However, an increase in $N_k$ will always result in an increase in $H_k$ for any value of $\wre$. 
\par
With this schematic picture of how the reheating era affects the inflationary energy scales, we will describe the results in the next section.
\section{\label{sec:sec3}Constraints on inflation and reheating}
\begin{figure}[t]
\centering
 \includegraphics[scale=1]{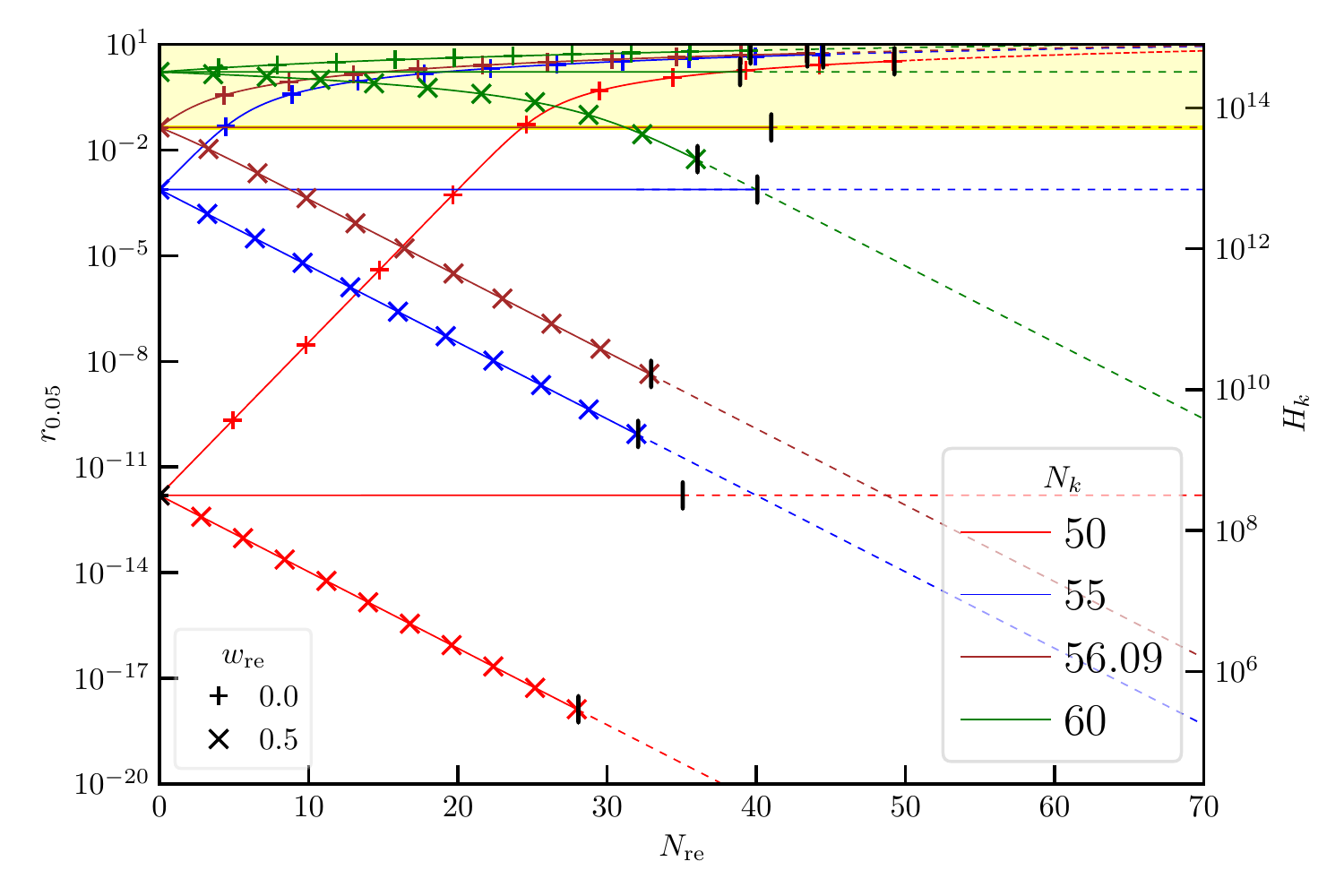}
 \caption{\scriptsize The Constrains on $r_k$ and $H_k$ from reheating. The red, blue, and green lines correspond to $N_k$ values $50,~55,$ and $60$ respectively, while the brown line is for the critical $N_k=56.09$, as described in the text. The horizontal lines are when the reheating is described by an effective EoS parameter $\wre=1/3$. The lines marked with `$+$' are for $\wre=0$ while lines with `$\times$' marks are for $\wre=0.5$. The vertical black bar on each line corresponds to the that reheating efolding $\nre$, when the temperature reaches $T_{\rm BBN}=10$ MeV. We have plotted the lines beyond this efolding as dotted lines. The shaded region is excluded from the upper bound on $r_k$ set by the Planck constraints on the tensor-to-scalar ratio~\cite{Tristram:2020wbi}.}
 \label{fig:result}
\end{figure}

We have seen how the presence of a phase of reheating affects the inflationary energy scale. We will now use the bound on tensor-to-scalar ratio and the bound on reheating temperature from BBN to constrain the reheating phase and inflationary epoch. We have seen that the qualitative behavior of $H_k$ and $r_k$ varies similarly depending on the parameter~$\wre$. We will use three values of $\wre$ viz. $\{0,1/3,0.5\}$ to illustrate our results. First, we will take $\wre=0$, i.e., we will consider that the effective fluid during reheating behaves as matter or pressure-less dust. Such a scenario is well-motivated when reheating occurs after inflation with scalar field potentials whose behavior around the minimum can be described as $V(\phi)\propto\phi^2$. In Fig~\ref{fig:result}, the variation of $r_k$ with reheating efolding number for three different values of $N_k$ are shown. Next, reheating with $\wre=1/3$, i.e., when the effective fluid is thermodynamically behaving as radiation, is a special case and can arise after inflation with quartic potentials. In the standard reheating constrain analysis, the reheating efolding number is indeterminate for $\wre=1/3$~\cite{Cook:2015vqa}. Physically this is not surprising as we are describing the reheating phase with a single thermodynamic quantity viz. its equation of the state parameter, so when it behaves as the radiation, it is impossible to distinguish reheating from the following radiation dominated phase. In the conventional scheme, this fact appears in the form of a singularity in the relation for $\nre$~\cite{Dai:2014jja}. However, we follow a different route in this work: keeping both the efoldings as independent variables and using the bounds on $\tre$ and $r_k$. We evaluate the bounds on the duration of reheating using (\ref{eq:res_rk}). Consequently, we can now have a quantitative measure on the reheating phase even when $\wre=1/3$. In Fig\ref{fig:result}, these are plotted as the horizontal lines for different $N_K$. The final representative value is chosen to be $\wre=0.5$; this will be the value of the EoS when reheating phase is due to coherently oscillating inflation in a $V(\phi)\propto\phi^6$ potential. It is also interesting that \cite{Figueroa:2019paj}, the authors found that the GW background remains detectable by LISA if the reheating EoS is within the narrow range $0.46 \lesssim w \lesssim 0.56$. In our previous work~\cite{Saha:2020bis}, on taking into effect the time variation of EoS during reheating from the epoch of coherent oscillation to the end of reheating, the effective EoS approaches a constant value $\wre\to0.44$ as $n\to\infty$. However, taking the whole period into the averaging, this value will increase.
\par
From the figure, we notice that when for $\nre=0$, i.e., when the reheating is instantaneous, $N_k=56.09$ saturates the bound on $r_k$ for any value of $\wre$. We have seen from the discussion in the last section that the inflationary energy scale $H_k$ (and consequently $r_k$) increases with $N_k$; thus, this value of $N_k$ turns out to be the maximum possible value of inflationary efolds admissible from the current bound on the inflationary tensor-to-scalar ratio from Planck and BICEP2~\cite{Tristram:2020wbi}. Consequently, as $H_K\propto\nre$ for fixed $N_k$ when $\wre<1/3$, if we have inflation with $N_k=56.09$, only instantaneous reheating is possible when reheating is due to non-stiff equation of state parameters~($\wre<1/3$). However, a finite reheating is possible for $\wre>1/3$, with the maximum reheating efolding set from the lower bound on $\tre$. For instance, when $\wre=0.5$ and we have inflation with this critical efolding number, the maximum reheating efolding is $\nre=32.48$ to satisfy the BBN bound on reheating temperature. For any inflationary efolding higher than this value, reheating can happen only with the system described by a stiff equation of state parameter, i.e., $\wre>1/3$. Furthermore, such exotic reheating with $N_k>56.09$ will have a lower bound on the duration of reheating, which is different from zero. For instance, when $N_k=60$ and $\wre=0.5$, the said two bounds forces reheating duration to be within $\{31.22,35.95\}$ excluding the possibility of instantaneous reheating.
\par
Another interesting feature in our present formulation, as we have described, is that we can now quantify the duration of reheating even if the effective fluid describing reheating behaves as radiation. As a case in point, for $N_k=55$, the reheating should not last for more than $34.96$ efolds so as not to violate the BBN bound on reheating temperature. While for $N_k=50$, the duration should not exceed $34.96$ efolds. Our formalism thus constrain the inflationary phase, i.e., the inflationary efolds and energy scale, and the duration reheating from the Planck bound on tensor-to-scalar ratio and BBN constrain on reheating temperature without referring to a specific model.
\section{\label{sec:conc}Discussion and conclusions}
The phase of reheating is crucial in the early universe. First, it replenishes the universe with matter and radiation so that the evolution of the standard radiation-dominated universe sets off. Additionally, the reheating constraints on inflationary models provide complementary information of the models~\cite{Dai:2014jja,Martin:2014nya,Munoz:2014eqa,Cook:2015vqa}. In this work, we have explored the avenue of incorporating the study of reheating constraints with the CMB observables from the inflationary era in a model-independent approach. Our main motivation was to constrain the quantities considering only slow-roll inflation so that we do not require dealing with a specific inflationary model. The basic assumption in our analysis is that inflation is due to a single canonical scalar field such that the inflationary energy scale~($H_k$) is related the amplitude of scalar~($A_s$) and the tensor-to-scalar ratio~($r_k$) as $H_k = \pi\Mp \sqrt{r_kA_s/2}$. It is worth mentioning that single field canonical inflation is currently the most favored by the observations~\cite{Akrami:2018odb}. Another crucial assumption is the relation in~(\ref{eq:hk_hend})  that is used to relate the inflationary energy scales at the end of inflation to that when the mode of interest leaves the horizon for slow-roll inflation. However, it can be easily extended to other cases with additional CMB observables to constrain the phase further, as we will show in subsequent work. The constraints that we have used are the bound on the inflationary tensor-to-scalar ratio ($r_k$) and the lower bound on the reheating temperature from BBN predictions. The summary of results is presented in Table~\ref{tab:summary_tab}.
\begin{table}
\begin{tabularx}{\textwidth}{c|XXcc}
\hline
\hline
$w_{\rm re}$ &$N_k$ 
& Allowed $N_{\rm re}$ & $r_k$ & $H_k~$(GeV)\\
\hline
$0$ & $56.09$ & $\{0\}$ & $\{0.044\}$ & $\{5.3\times10^{13}\}$\\
\cline{2-5}
 & $55$ & $\{0,4.36\}$ & $\{0.00075,0.044\}$ & $\{6.9\times10^{12},5.3\times10^{13}\}$\\
 \cline{2-5}
  & $50$ & $\{0,24.3\}$ & $\{1.6\times10^{-12},0.044\}$ & $\{3.16\times10^{8},5.3\times10^{13}\}$\\
\hline
$1/3$ & $55$ & $\{0,39.95\}$ & $\{0.00075,0.044\}$ & $\{6.9\times10^{12},5.3\times10^{13}\}$\\
 \cline{2-5}
  & $50$ & $\{0,34.96\}$ & $\{1.6\times10^{-12},0.044\}$ & $\{3.16\times10^{8},5.3\times10^{13}\}$\\
  \hline
$0.5$  & $60$ & $\{31.22,35.95\}$ & $\{5.5\times10^{-3},4.4\times10^{-2}\}$ & $\{1.9\times10^{13},5.3\times10^{13}\}$\\
  \cline{2-5}
 & $56.09$ & $\{0,32.84\}$ & $\{0.044,4.4\times10^{-9}\}$ & $\{5.3\times10^{13},1.7\times10^{10}\}$\\
\cline{2-5}
 & $55$ & $\{0,31.97\}$ & $\{0.00075,8.6\times10^{-11}\}$ & $\{6.9\times10^{12},2.4\times10^{9}\}$\\
 \cline{2-5}
  & $50$ & $\{0,27.3\}$ & $\{1.6\times10^{-12},1.3\times10^{-18}\}$ & $\{3.16\times10^{8},2.9\times10^{5}\}$\\
\hline
\hline
\end{tabularx}
\caption{Summary of results.}
\label{tab:summary_tab}
\end{table}
With these simple assumptions, we have seen that reheating constraints, namely the requirement of a successful BBN and the upper bound on $r_k$, have significantly constrained the reheating efolding number in an entirely model-independent approach. In this way, we have constrained not the inflationary models but the inflationary phase itself.
\par
Finally, considering the instantaneous reheating~($\nre=0$) as a benchmark, we see from our analysis that if the pivot scale exits the horizon at $50$ efolds before the end of inflation~(i.e., $N_k=50$), the value of tensor-to-scalar ratio is found to be $r_k=1.6\times10^{-12}$ that indicates a low scale inflation~($H_k=3.1\times10^8$GeV). In situations like this, a signal of inflationary stochastic gravitational waves on the CMB scale is unlikely to be detected in the GWs missions. However, for higher values of $N_k$, the inflationary gravitational wave signal can, in principle, be detected in future missions~\cite{Hu:2017yoc,Baker:2019pnp,Bailes:2021tot}. Moreover, in all such cases, a finite amount of reheating will indicate a higher value of inflationary energy scale and, consequently, higher $r_k$, thus improving the detection possibility. However, we should also note that the upper bound on $r_k$ set an upper bound on the inflationary efolding number as $N_k=56.09$. Hence if the inflationary efolding number corresponds to this maximum value, then canonical reheating with $\wre<1/3$ should be instantaneous. On the other hand, if reheating is due to some exotic component ($\wre>1/3$), a finite reheating will decrease the inflationary scale and reduce the detection possibility.
\begin{acknowledgments}
I want to thank Prof. L Sriramkumar for his comments on the manuscript. 
I wish to acknowledge support from the Science and Engineering Research Board, Department of Science and Technology, Government of India, through the Core Research Grant CRG/2018/002200.
\end{acknowledgments}
\appendix
\section{\label{app:appA} Model dependence in \texorpdfstring{$H_{\rm end}/H_{k}$}{HkHend}}
When relating the Hubble parameters at the end of inflation to that when a scale $k$ leaves the Hubble horizon, we have used the fact that the integration in (\ref{eq:int_hk}) gets most of its contribution when the field is around $\phi_k$. Such an approximation is justified for single field slow-roll inflation where the potentials have a large plateau in their field space. Although such models are currently favored from observables, we must examine how such an approximation fares with some conventional inflationary models. A model dependence may occur even in plateau potentials if there is a violation of the slow-roll approximation for a brief period as customary with the models of ultra slow-roll and punctuated inflation~\cite{Allahverdi:2006iq} or if the potential has a step in the field space~\cite{Adams:2001vc}. In such instances, we can expect a deviation from the above simple relation.
All such cases can be conveniently incorporated by defining a \textit{flatness parameter} $\delta$ such that:
\begin{equation}
\frac{H_{\rm end}}{H_k} = \delta \exp{-\epsilon_k N_k}
\end{equation}
For a very flat potential, $\delta=1$. 
\begin{figure}[!ht]
    \centering
    \includegraphics[width=\textwidth]{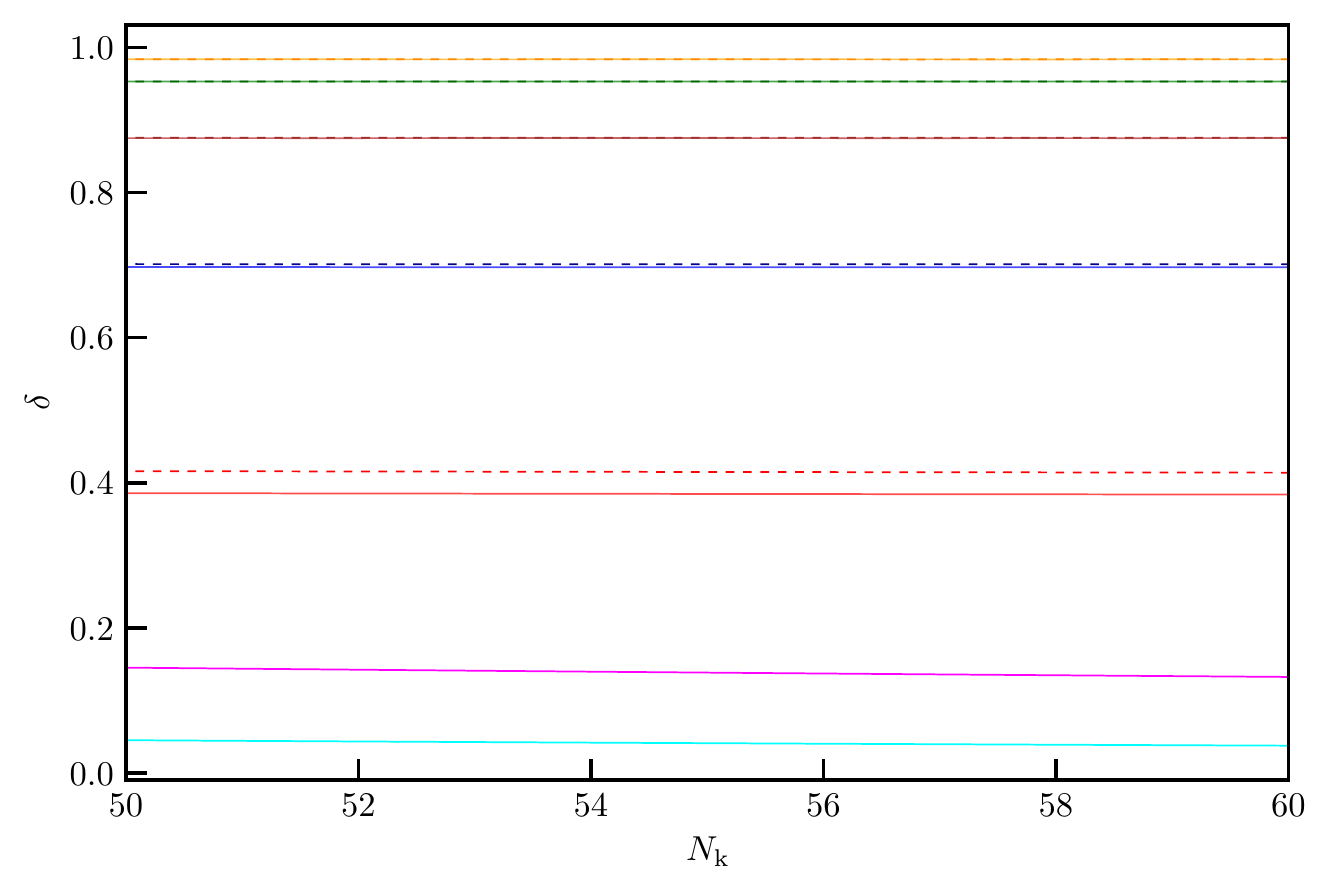}
    \caption{Plot of the quantity $\delta$ which measure the accuracy of the relation defined in (\ref{eq:hk_hend_0}) to approximate the ratio of $H_{\rm end}$ to $H_k$. Apart from the large field models: the $\lambda\phi^4$~(cyan) and the $m^2\phi^2$~(magenta) models, we have plotted $\alpha$-attractor T-model as the representative of the plateau-type potentials. The plot legends are as follows: solid line for $n=4$ and dashed are for $n=2$ with the values of $\alpha$ to be $\alpha=1$~(red), $\alpha=10^{-1}$~(blue), $\alpha=10^{-2}$~(brown), $\alpha=10^{-3}$~(green) and $\alpha=10^{-4}$~(orange). As we decrease $\alpha$, the potential gets flatter thereby improving the approximation.}
    \label{fig:figApp}
\end{figure}
To study the model dependence, we will solve the inflationary dynamics equations to determine $\delta$. The results are shown in Fig (\ref{fig:figApp}). We see that for large field models such as chaotic $\lambda\phi^4$ inflation $\delta\sim0.14$, while for $m^2\phi^2$ inflation it is around $0.2$. As we mentioned before, this approximation gets better as the potential gets more and more flat, as in the case of plateau-type potentials. Solving for the $\alpha$-attractor T-model~\cite{Kallosh:2013hoa} given by the potential $V(\phi)=\Lambda^4\tanh\left(\frac{1}{\sqrt{6\alpha}}\frac{\phi}{\Mp}\right)^{n}$. We find that, as we decrease the value of $\alpha$~(the parameter that controls the flatness of the potential), the approximation becomes more and more accurate. Nonetheless, as evident from Eq. (\ref{eq:hk_hend}), the contribution of $\delta$ will be logarithmic and can be neglected for large $N_k$. A more general treatment will require considering higher-order slow-roll parameters and eventually require constraints on additional CMB observables. We left those details for a subsequent publication.
\clearpage
\providecommand{\href}[2]{#2}\begingroup\raggedright\endgroup

\end{document}